\documentclass[runningheads]{llncs}

\usepackage{geometry}
\usepackage[parfill]{parskip}
\usepackage{graphicx}
\usepackage{amssymb}
\usepackage{epstopdf}
\usepackage[utf8]{inputenc}
\usepackage[english]{babel}
\usepackage[T1]{fontenc}
\usepackage{subcaption}
\usepackage{amsmath}
\usepackage{proof}
\usepackage[]{algorithm2e}
\usepackage{bm}
\usepackage{listings}
\usepackage{fp}
\usepackage{xcolor}
\usepackage{colortbl}
\usepackage{booktabs}
\usepackage{multirow}
\usepackage{enumitem}
\usepackage{multicol}
\usepackage{footmisc}
\usepackage[normalem]{ulem}

\newcommand{\souffle}{Souffl\'{e}~}

\newcommand{\FM}{\mathcal{FM}}
\newcommand{\featset}{\mathcal{F}}

\newcommand{\conf}{\rho}

\newcommand{\term}[1]	{\emph{#1}}

\newcommand{\true}  {True~}
\newcommand{\false}  {False~}

\begin{document}
    
\title{Variability-Aware Datalog}
\author{Ramy Shahin \and Marsha Chechik}
\institute{University of Toronto \\
           \email{\{rshahin, chechik\}@cs.toronto.edu}}
\maketitle

\begin{abstract}
Variability-aware computing is the efficient application of programs to different sets of inputs that exhibit some variability. 
One example is program analyses applied to Software Product Lines (SPLs). 
In this paper we present the design and development of a variability-aware version of the \souffle Datalog engine. 
The engine can take facts annotated with Presence Conditions (PCs) as input, and compute the PCs of its inferred facts, eliminating facts that do not exist in any valid configuration. 
We evaluate our variability-aware \souffle implementation on several fact sets annotated with PCs to measure the associated overhead in terms of processing time and database size.

\end{abstract}

\newcommand{\ort}{\textbackslash/~}
\newcommand{\andt}{/\textbackslash~}
\section{Introduction}
\label{sec:motivation}
A Datalog engine is used to infer knowledge from a set of facts given some inference rules. There are cases though where we need to apply the same rules to different sets of facts coming from different worlds, or different configurations. For example, the Doop~\cite{Bravenboer:2009} pointer analysis framework encodes its logic as Datalog rules, and applies them to facts extracted from Java programs. Doop can only work on a single software product at a time. However, it is common for software engineers to develop a whole family of products, a Software Product Line (SPL)~\cite{Clements:2001}, as one project, exploiting the commonality across those products. Different \term{variants} (products) implement different sets of \term{features}. Since each feature can be either present or not in a variant, the number of variants is usually exponential in the number of features.

To use a framework like Doop on an SPL, we need to apply it to each of the variants individually. This is infeasible in most cases because of the exponential number of variants. Also, it involves a lot of redundancy because it does not leverage the commonality across variants.
To mitigate those drawbacks, some program analyses have been \term{lifted} to efficiently work on SPLs instead of single products~\cite{Bodden:2013,Classen:2010,Gazzillo:2012,Kastner:2012,Midtgaard:2015,Salay:2014,Thum:2014}. This lifting process usually invovles reimplementing the analysis to be variability-aware.

Our prior work~\cite{Shahin:2019} outlines an approach to apply Doop (and similar frameworks) to the whole SPL at once, showing orders of magnitude of savings in computation time and storage space compared to running on each variant separately. One building block of that work was modifying the \souffle\cite{Jordan:2016} Datalog engine to be variability-aware, i.e., taking fact variability into consideration when inferring new facts.
One fundamental advantage of our approach is that lifting a Datalog engine to be variability-aware automatically lifts all analyses that use it. In addition, variability-aware inference can be widely applied beyond program analysis. In any application domain, it is possible for different facts to be present only in specific situations, configurations, or in some constrained worlds. Instead of modeling each of those variants separately, it makes sense to model them together since inference rules are orthogonal to variability. 

The rest of this paper starts with some background definitions and a motivating example (Sec.~\ref{sec:background}), followed by the design of variability-aware \souffle (Sec.~\ref{sec:implementation}). We then present the results of our evaluation experiments (Sec.~\ref{sec:evaluation}), and finally conclude and suggest some future directions (Sec.~\ref{sec:conclusion}).

\section{Background and Motivating Example}
\label{sec:background}

In this section we define some Datalog and variability terms, illustrating them on the motivating example in Fig.~\ref{fig:example}. We then briefly introduce the architecture of the \souffle Datalog engine.

\lstset{moredelim=[is][\sout]{|}{|}}

\newcommand{\A}{\text{Air}}
\renewcommand{\L}{\text{Land}}
\renewcommand{\S}{\text{Sea}}


\begin{figure*}[t]
\begin{subfigure}{0.49\textwidth}
\begin{lstlisting}[caption=Path rules., captionpos=b, label=lst:path]
Path(v1, v2) :- Edge(v1, v2).
Path(v1, v3) :- 
	Edge(v1, v2), Path(v2, v3).
\end{lstlisting} 
\begin{lstlisting}[caption=Variability-aware inputs., captionpos=b, label=lst:inputs]
Edge(Athens, Rome)   @ Sea.
Edge(Rome, Toronto)  @ Air.
Edge(NYC, Athens)    @ !Land.
Edge(Toronto, NYC)   @ Land.
\end{lstlisting} 
\end{subfigure}
~
\begin{subfigure}{0.49\textwidth}
\begin{lstlisting}[caption=Variability-aware outputs., captionpos=b, label=lst:outputs]
Path(Athens, Rome)    @ Sea.
Path(Rome, Toronto)   @ Air.
Path(NYC, Athens)     @ !Land.
Path(Toronto, NYC)    @ Land.
|Path(Athens, Toronto) @ Sea /\ Air.|
|Path(Rome, NYC)       @ Air /\ Land.|
|Path(Toronto, Athens) @ Land /\ !Land.|
Path(NYC, Rome)       @ Sea.
\end{lstlisting}
\end{subfigure}
\vspace{-0.2in}
\caption{Motivating example.}
\label{fig:example}
\vspace{-0.2in}
\end{figure*}

\subsection{Datalog and Variability}
\term{Datalog} is a declarative data definition and query language that combines relational data manipulation and logical inference~\cite{Ceri:1989}. A \term{Datalog program} is a set of inference rules, collectively referred to as the \term{Intentional Dataabse (IDB)}. For example, the Datalog program in Listing~\ref{lst:path} computes directed paths given graph edges.

A program takes \term{facts}, referred to as the \term{Extensional Database (EDB)}, as input, and by repeatedly applying the inferrence rules to the input facts new output facts are generated. Listings~\ref{lst:inputs} and~\ref{lst:outputs} are examples of input and output facts respectively.

\term{Variability-aware computing} is the ability to efficiently compute over values from different worlds at the same time. A set of worlds is defined in terms of a set of \term{features} $\featset$. A world is defined by a \term{configuration} $\conf$, where each feature can be either present or absent. A set of worlds is defined by a propositional formula over features. 

Each software artifact can be labeled with a \term{Presence Condition (PC)}: a propositional formula specifying the set of worlds in which this artifact exists. Datalog facts are an example of artifacts. If we are modeling a set of worlds defined by three features: \term{Land}, \term{Air} and \term{Sea}, facts can be labeled by PCs as seen in Listing~\ref{lst:inputs}. The '@' symbol is syntactically used to separate the fact predicate from its PC. We use the symbols '!' for negation, '\ort' for disjunction, and '\andt' for conjunction. Parenthesis can be also used to override operator precedence.

Usually not all feature combinations are valid. For example, the expression $(\text{Land} \land \text{Sea})$ states that we have an edge that is \term{both} overland and marine, which does not make sense. To rule out invalid feature combinations, a product line usually has a \term{feature model} $\FM$: a propositional formula over features specifying their valid combinations (valid worlds). A configuration $\conf$ is valid only if $\conf \land \FM$ is satisfiable. Our example's feature model is
\[
(\A \lor \L \lor \S) \land \neg(\A \land \L) \land \neg(\L \land \S) \land \neg(\S \land \A)
\] 

\vspace{-0.1in}

Now a \term{variability-aware Datalog engine} needs to take both the feature model and the presence conditions of facts into consideration when inferring new facts. Whenever a new fact is inferred, its Presence Condition (PC) should be the conjunction of the PCs of its resolvent facts together with the feature model. If this PC is not satisfiable, the inferred fact does not belong to any valid configuration (world), and can be removed.

Listing~\ref{lst:outputs} shows the results of applying our variability-aware Datalog engine to the program and facts aforementioned. Crossed-out facts are the ones removed because their presence conditions are not satisfiable (in general or with respect to the feature model). 

Formal syntax and semantics of variability-aware Datalog, together with correctness criteria of the lifted inference algorithm, and proof of correctness are presented in~\cite{Shahin:2019}. 



\subsection{\souffle}

\begin{figure*}[t]
	\centering
	\includegraphics[width=0.8\textwidth]{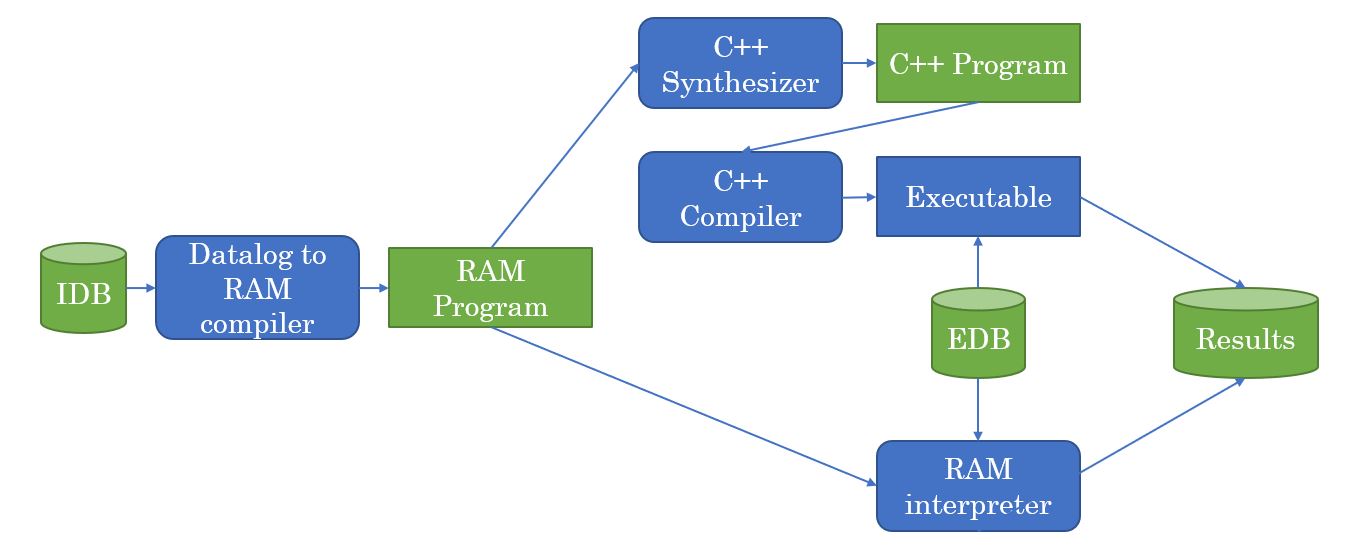}
	\caption{\souffle architecture.}
	\label{fig:souffle-arch}
	\vspace{-0.1in}
\end{figure*}

\souffle\cite{Jordan:2016} is an optimized Datalog engine, with a Datalog interpreter in addition to the option of compiling programs into native C++ code (Fig.~\ref{fig:souffle-arch}). \souffle first compiles Datalog into Relational Algebra Machine (RAM) programs, which are then either interpreted or compiled. RAM is a relational algebra language with a fixpoint operator.

\souffle employs a semi-naive Datalog evaluation algorithm to compile Datalog into RAM. Elaborate data indexing techniques and multi-threaded query processing are then used to evaluate RAM programs. These techniques, in addition to the ability to compile RAM into C++, and subsequently into optimized native machine code, result in high-performance exeuction of Datalog programs.

\section{Variability-aware \souffle}
\label{sec:implementation}



We modified the \souffle engine to support variability-aware Datalog inference. \souffle runs in two modes: interpreter mode and compilation (code synthesis) mode. We only support the interpreter mode at this time.

\subsection{Syntax Extension}
\begin{figure*}[t]
\centering
$
\begin{array}{lcl}
    \text{PC}  & ::= & \text{ID | !PC | (PC) | PC \ort PC | PC \andt PC} \\
    \text{ATOMLIST} & ::= & \text{ATOM | ATOM, ATOMLIST}  \\
    \text{FACT} & ::= & \text{ATOM . | ATOM @ PC .} \\
    \text{RULE} & ::= & \text{ATOM :- ATOMLIST .}
\end{array}
$
\caption{BNF syntax of \souffle clauses and presence conditions.}
\label{fig:pc-syntax} 
\end{figure*}

We extend the \souffle fact syntax (Fig.~\ref{fig:pc-syntax}) with an optional \term{Presence Condition (PC)} before the period ('.') at the end. A presence condition is prefixed with the '@' symbol, and has the syntactic structure of a propositional formula.

\begin{figure*}[t]
	\centering
    \includegraphics[width=0.8\textwidth]{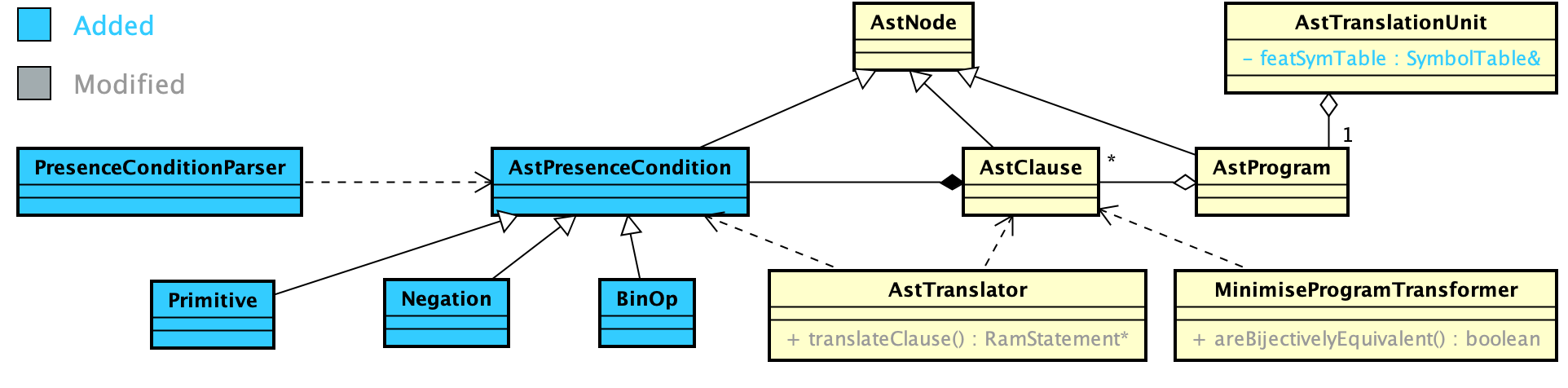}
    \caption{Modifications and additions to \souffle syntax and parsing classes.}
    \label{fig:souffle-syntax}
    \vspace{-0.1in}
\end{figure*}

The \souffle grammar (Lex and Yacc files) is extended accordingly, and Abstract Syntax Tree (AST) classes are added to the code-base for Presence Conditions (Fig.~\ref{fig:souffle-syntax}). \term{AstPresenceCondition} is an abstract class inheriting from \term{AstNode}. Concrete subclasses of \term{AstPresenceCondition} are \term{Primitive} (for \true, \false and atomic propositional symbols), \term{Negation}, and \term{BinOp} (for conjunction and disjunction).
    
The syntactic category of presence conditions can appear in \souffle programs, and also in CSV files. While the \souffle parser takes care of programs, we had to implement a separate parser for PCs appearing in CSV files (the \term{PresenceConditionParser} class). It identifies a PC as an optional field prefixed with '@' coming at the end of a fact. If a PC exists, it is parsed into an \term{AstPresenceCondition} object.

The \term{AstClause} class is now extended with an \term{AstPresenceCondition} field. Unless a PC is provided for a clause, the default value is the \term{\true}proposition (indicating that the fact is present in all configurations). \term{AstTranslator} has a method called \term{translateClause} that compiles an \term{AstClause} into a \term{RamStatement}. This method is modified to translate the PC of the clause as well.
  
The propositional symbols used in PCs come from a syntactic category different from that of \souffle variables and constants. To avoid name collisions, we store those symbols in a separate symbol table (\term{featSymTable}). An \term{AstTranslationUnit} now has two symbols tables: one for Datalog symbols and the other for propositional symbols (feature names).
    
\souffle performs some optimizations on the AST before it is translated into a RAM program. For example, in the \term{MinimiseProgramTransformer} class, \term{areBijectivelyEquivalent} is a method that checks if two clauses are bijectively equivalent. We extend this method to compare the PCs of the clauses as well. If the PCs are not syntactically the same, we consider the two clauses not equivalent.

\subsection{RAM}
\begin{figure*}[t]
	\centering
    \includegraphics[width=0.8\textwidth]{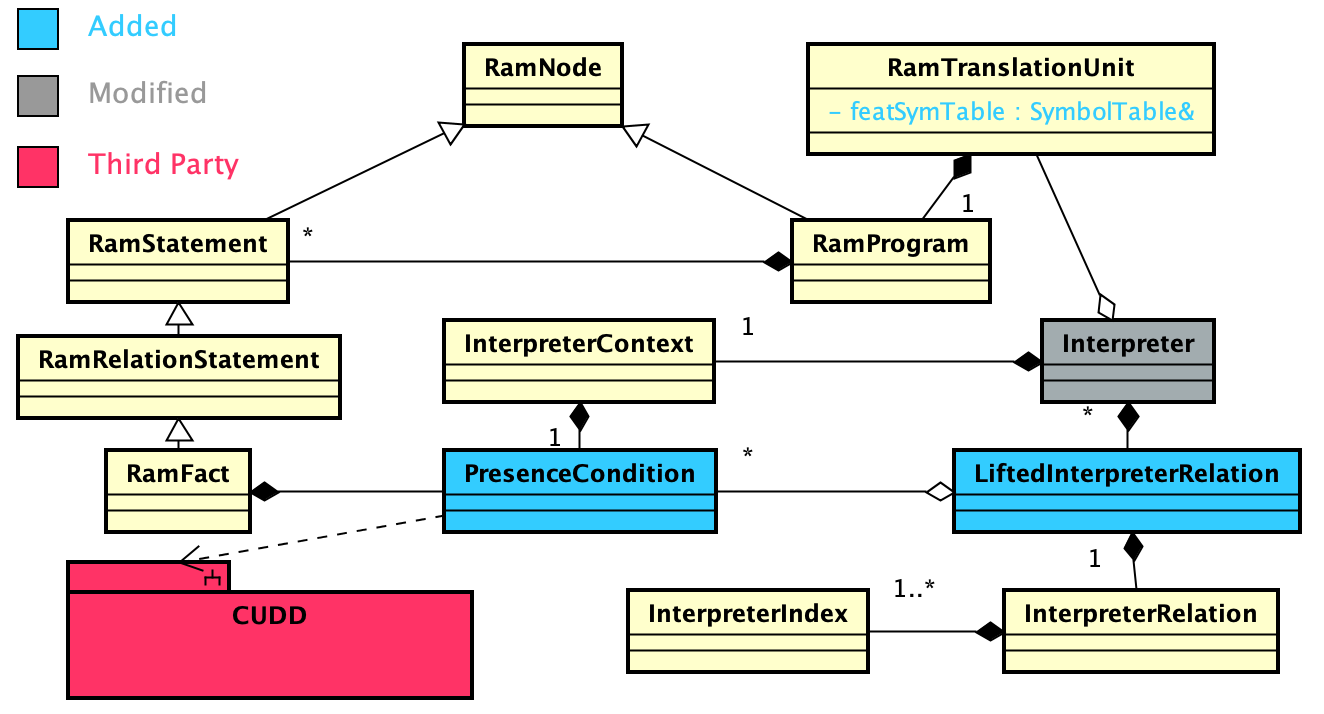}
    \caption{Modifications and additions to RAM interpreter classes.}
    \label{fig:ram}
    \vspace{-0.1in}
\end{figure*}

The AST of a \souffle translation unit is compiled into a \term{Relational Algebra Machine (RAM)} program, encapsulated in a \term{RamTranslationUnit} object (Fig.~\ref{fig:ram}). Similar to \term{AstTranslationUnit}, we need to carry the feature symbol table (\term{featSymTable}) over to RAM as a part of the translation unit. A \term{RamProgram} is contained within a translation unit, and it consists of a set of \term{RamStatement} objects. A \term{RamFact} is a special kind of \term{RamStatement}, and we add a \term{PresenceCondition} object as a field to it.

A syntactic \term{AstPresenceCondition} is compiled into a \term{PresenceCondition} object, which encapsulates a representation of the PC propositional formula. We store PCs as Binary Decision Diagrams~\cite{Huth:2004}, and we use CUDD~\cite{Somenzi:1998} as a BDD engine. To keep the number of PC objects at a minimum, we also maintain a hash-table mapping BDDs to PC objects. This way a new PC object is created only if no other object with the same BDD already exists in memory.
 
\souffle stores RAM relations as tables of numbers. String values are stored elsewhere, and their corresponding numeric identifiers are the values actually stored in relations. This keeps relations homogeneous, easy to access and index. Since we now need to add a PC for each RAM record, the easiest way is to extend relations with an extra field for the PC. To keep the relation data-structure homogeneous, instead of storing a PC object, we store its address, which is a 64-bit numeric value, pretty much like other fields. This way our extra PC field is opaque to the rest of the RAM subsystem. We had to take special care of \term{nullary relations}, i.e., relations of zero fields. They have special semantics in Souffl\'{e}, and to preserve the semantics, we consider a relation of a single field (the PC) to be nullary.

\subsection{Interpreter}

The \souffle interpreter runs a program on the fly, keeping a context of type \term{InterpreterContext}, and manipulating a set of RAM relations. To avoid getting into the details of how relations are stored, and how data indices are maintained, we decided not to modify \term{InterpreterRelation} and \term{InterpreterIndex}. Instead, we wrap \term{InterpreterRelation} in \term{LiftedInterpreterRelation}. The wrapper maintains the same interface, but adds the semantic manipulation of the PC field. 

Another significant difference between \term{LiftedInterpreterRelation} and \term{InterpreterRelation} is existence checking of records. In \souffle checking if a record exists in a relation is straightforward using the full index of the relation, returning true if the record exists in the index and false otherwise. With PCs existence checking is more subtle because the record we are looking for might exist but with a different PC. To accommodate for this, we add a PC output parameter to \term{exists}, the existence checking method of \term{LiftedInterpreterRelation}. Now instead of just returning a boolean indicating whether a record exists in a relation, we also return a pointer to the stored PC of the record (if the record exists).

Now whenever two records are resolved by the interpreter, their PCs need to be conjoined, and the conjunction (if satisfiable) becomes the PC of the resulting record. If on the other hand the conjunction is not satisfiable, the result can be safely ignored because an unsatisfiable PC indicates an empty set of configurations in which this record exists. Satisfiability checking is a constant-time operation on BDDs (although BDD construction might take exponential time in the number of variables). Because clause resolution might take place recursively, we add a PC field to \term{InterpreterContext}, which keeps track of the PCs of intermediate results.  

When inserting a record into a relation, again we need to take the PC into consideration. If that record already exists in the relation with the same PC, then we do not need to add it again. If on the other hand it exists with a different PC, we now need to disjoin that with the new PC because we are expanding the set of configurations where this record exists into that of the union of the two PCs. If the record does not exist at all, we just add it with its new PC.


We had to modify the I/O subsystem of \souffle to make sure we correctly read and write PCs together with records from/to CSV files. \term{PresenceConditionParser} is used to parse PCs on input, and logic for serializing PCs is added to the \term{PresenceCondition} class. At this point, we do not support storing facts to SQLite databases.

\section{Evaluation}
\label{sec:evaluation}

\begin{figure*}[t!]
	\centering
\begin{subfigure}[t]{0.49\textwidth}
	\centering
	\includegraphics[width=\textwidth]{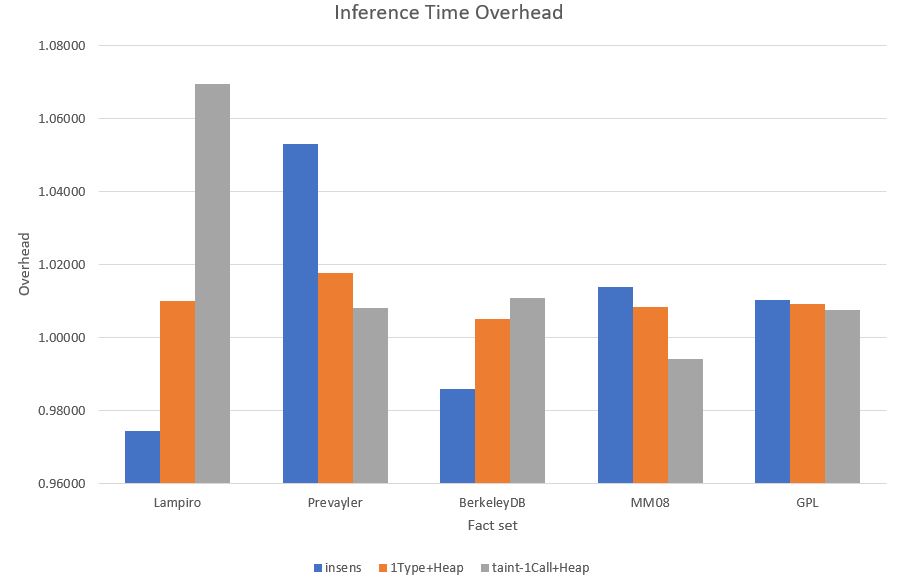}
	\caption{Time overhead.}
	\label{fig:time}
\end{subfigure}
~
\begin{subfigure}[t]{0.49\textwidth}
	\centering
	\includegraphics[width=\textwidth]{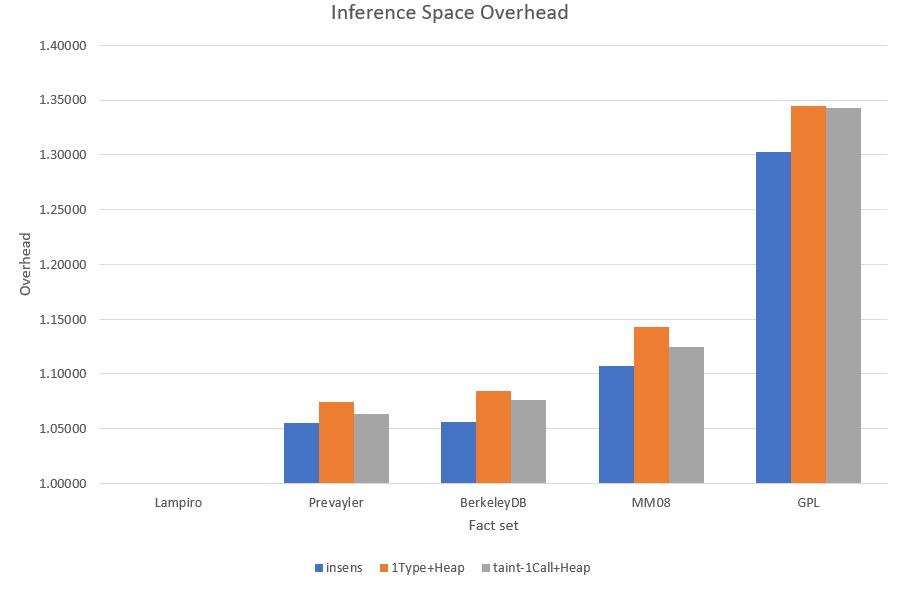}
	\caption{Space overhead.}
	\label{fig:space}
\end{subfigure}
\caption{Time and space overhead due to variability-aware inference for five different fact sets and three sets of rules.}
\vspace{-0.1in}
\end{figure*}

\begin{table*}[tbp] 
  \centering
  \caption{Inference time for three different Datalog programs applied to five different fact sets. For each fact base we report the number of features (R), number of facts with PCs other than True (FPC), inference time (T), database size (S), non-variabality-aware inference time (TN), and non-variability-aware database size (SN). Time is reported in milliseconds, and space is reported in Kilobytes.}
  \scriptsize
    \begin{tabular}{lrrrrrrrrrrrrrr}
    	\toprule
        \multicolumn{3}{l}{} & \multicolumn{4}{c}{insens} &        
        \multicolumn{4}{c}{1Type+Heap} & 
        \multicolumn{4}{c}{taint-1Call+Heap} \\
    \cmidrule(r){4-15}
    Fact-base & \multicolumn{1}{l}{R} & \multicolumn{1}{l}{FPC} & \multicolumn{1}{l}{T(ms)} & \multicolumn{1}{l}{S(KB)} & \multicolumn{1}{l}{TN(ms)} & \multicolumn{1}{l}{SN(KB)} & \multicolumn{1}{l}{T(ms)} & \multicolumn{1}{l}{S(KB)} & \multicolumn{1}{l}{TN(ms)} & \multicolumn{1}{l}{SN(KB)} & \multicolumn{1}{l}{T(ms)} & \multicolumn{1}{l}{S(KB)} & \multicolumn{1}{l}{TN(ms)} & \multicolumn{1}{l}{SN(KB)} \\
    \midrule
    Lampiro & 18 & 343 & 8,111 & 41,170 & 8,324 & 41,160 & 20,725 & 149,686 & 20,522 & 149,661 & 45,996 & 230,370 & 43,014 & 230,329 \\
    Prevayler & 5 & 6,507 & 5,334  & 4,407 & 5,066  & 4,177 & 6,013  & 8,630 & 5,908  & 8,035 & 9,717  & 5,534 & 9,640  & 5,203 \\
    BerkeleyDB & 42 & 49,062 & 10,810 & 49,725 & 10,966 & 47,071 & 17,273 & 122,922 & 17,186 & 113,346 & 21,474 & 112,060 & 21,247 & 104,137 \\
    MM08  & 27 & 6,811 & 4,720  & 3,259 & 4,656 & 2,944 & 5,142  & 6,990 & 5,099  & 6,114 & 9,306  & 7,829 & 9,360  & 6,960 \\
    GPL   & 21 & 3,353 & 4,517  & 409   & 4,471  & 314 & 4,718  & 593   & 4,675  & 441   & 8,861  & 462   & 8,795  & 344  \\
    \bottomrule
    \end{tabular}%
  \label{tab:results}%
  \vspace{-0.1in}
\end{table*}%

We evaluate the performance of our implementation of variability-aware \souffle in terms of time and space overhead. In particular, the research question we are trying to answer is how much of an overhead in terms of inference time and database size is attributed to our modifications to \souffle. To answer this question, we compare the performance of \souffle on a fact set annotated with PCs against its performance on the same set with the PCs removed.

We use the same dataset used in~\cite{Shahin:2019}, which is comprised of five fact sets extracted from Java programs, and three program analyses (implemented as Datalog rules) applied to each of them. Table~\ref{tab:results} summarizes the number of features (R) and number of facts annotated with PCs (FPC) for each of the five benchmark fact sets. In addition, for each of the three Datalog rule sets (insens, 1Type+Heap, taint-1Call+Heap) it outlines the inference time (T), database size after inference (S), and the corresponding values when the fact set with no PC annotations is used (TN and SN respectively). Time is measured in milliseconds, and space is measured in Kilobytes.

Fig.~\ref{fig:time} shows the inference time overhead when applying each of the three Datalog programs to each of the five fact sets. Overhead is calculated as a ratio between the time taken by variability-aware inference to standard Datalog inference. There are a few cases of overhead values less than 1.0, which can be considered as outliers due to other factors affecting overall processing time (e.g., I/O). From this graph, we can conclude that the overhead is relatively small (7\% was the maximum reported for taint-1Call+Heap on Lampiro). We still can not see a direct correlation between the time overhead and fact set attributes (e.g., feature count, percentage of facts annotated with PCs).

Similarly, Fig.~\ref{fig:space} shows the database size overhead when applying the same Datalog programs to the fact sets, where the ratio here is between database sizes. \souffle databases are stored as text files, and since variability-aware facts (including inferred ones) might have PCs, and those PCs are stored as text, it is natural that a variability-aware fact database takes more space than a plain databse with no PCs. We can see from this graph that the database size overhead grows roughly with the percentage of PC-annotated input facts. This overhead reaches almost 34\% for GPL, where about 60\% of the input facts are PC-annotated.

Please recall that the rationale behind variability-aware computing is to run a program only once on values from all configurations, as opposed to running the program on each configuration separately. Since the number of configurations is typically exponential in the number of features, the marginal overhead we see here is negligible compared to the savings due to running the program only once. More details on our experiment setup and evaluation results can be found in~\cite{Shahin:2019}.

\section{Conclusion and Future Work}
\label{sec:conclusion}

In this paper we presented the design and development of the variability-aware \souffle Datalog engine. The engine can take Datalog facts annotated with presence conditions as input, and compute the presence conditions of its inferred facts, eliminating facts that do not exist in any valid configuration.

We evaluated the overhead of our variability-aware Datalog inference in terms of inference time and size of the fact database, showing that time overhead is marginal, and space overhead grows with the percentage of PC-annotated input facts. This overhead is acceptable compared to the brute force approach (each configuration running separately), where the number of configurations, and accordingly the overhead, is exponential in the number of variability features.

For future work, we plan to extend our variability-aware inference implementation to the \souffle C++ code generator. We also plan to extend our theoretical foundations and implementation to support presence conditions on rules. This would allow for variability of inference logic in addition to data. 

\bibliographystyle{splncs04}
\bibliography{spl,datalog}
\end{document}